# Ferroelectricity induced by oxygen vacancies in relaxors with perovskite structure

**Maya D. Glinchuk[1], Eugene A. Eliseev[1], Guorong Li[2], Jiangtao Zeng[2], Sergei V. Kalinin[3] and Anna N. Morozovska[4,*]**

[1] *Institute for Problems of Materials Science, National Academy of Sciences of Ukraine, Krjijanovskogo 3, 03142 Kyiv, Ukraine*

[2] *Key Laboratory of Inorganic Functional Material and Device, Shanghai Institute of Ceramics, Chinese Academy of Sciences, 1295 Dingxi Road, Shanghai 200050, China*

[3] *The Center for Nanophase Materials Sciences, Oak Ridge National Laboratory, Oak Ridge, TN 37831*

[4] *Institute of Physics, National Academy of Sciences of Ukraine, 46, pr. Nauky, 03028 Kyiv, Ukraine*

## Abstract

The consideration of oxygen vacancies influence on the relaxors with perovskite structure was considered in the framework of Landau-Ginzburg-Devonshire phenomenological theory. Main attention was paid to PZN-PLZT relaxor, where earlier experimental investigation of oxygen vacancies influence on the polar properties was performed and the evidence of oxygen vacancies induced ferroelectricity was obtained.

Since the oxygen vacancies are known to be elastic dipoles, they influence upon elastic and electric fields due to Vegard and flexoelectric couplings. We have shown that a negative Curie temperature $T_C^*$ of a relaxor is renormalized by the elastic dipoles due to the electrostriction coupling and could become positive at some large enough concentration of the vacancies. Positive renormalized temperature $T_C^R = T_C^* + \Delta T$ is characteristic for ferroelectric state. At $T < T_C^R$ all the polar properties could be calculated in the way conventional for ferroelectrics, but obtained experimental data speaks in favor about the coexistence of ferroelectric phase with relaxor state, i.e. about the morphotropic region in PZN-PLZT relaxor. At $T > T_C^R$ random field characteristic for relaxors conserves, but since the mean square deviation of polarization is nonzero the coexistence with dipole glass state is not excluded. For the case $T > T_C^R$ we calculated the local polarization and electric field induced by the flexo-chemical coupling with oxygen vacancies.

---

[*] Corresponding author: anna.n.morozovska@gmail.com

# 1. INTRODUCTION

The broad applications of relaxor ferroelectrics in the modern sensors, actuators, high performance electromechanical transducers and other electronic devices [1, 2] generates the permanent interest to investigation and fabrication of these materials. These applications are based on peculiar properties of the relaxors absent in ordinary ferroelectrics [3]. The peculiarities appear due to random electric fields induced by two factors. The first one is substitutional disorder in cations positions because general formula can be written as $A_{1-x}A'_xB_{1-y}B'_yO_3$, which leads to the local shift of ions from their conventional equilibrium positions. The second one is the presence of vacancies and other unavoidable defects. In previous years the influence of random field was considered in many details (see e.g. references in [3] with special attention to [4, 5]), but the attention was paid mainly to the first factor inherent to any relaxor. Allowing for fact that the concentration of e.g. oxygen vacancies ($V_O$) can be changed, it could open the way to govern relaxor properties on demand under the condition of the influence of $V_O$ on relaxor properties to be studied. To the best of our knowledge **$V_O$** were considered theoretically mainly as random field sources up to now. As to the experimental papers we would like to draw attention to paper [6], where influence of $V_O$ on phase diagram and properties were studied. The authors considered PZN-PLZT relaxor. They increase the $V_O$ concentration by addition of nitrogen flow when sintering the relaxor (**NS samples**) and after this procedure some of the samples were annealed in oxygen (**OA samples**). Comparative analysis of dielectric permittivity temperature dependence of NS and OA samples had shown that relaxor characteristics were suppressed by inducing of oxygen vacancies with high concentration. In other words $V_O$ added ferroelectric (**FE**) phase to the relaxor. The aim of this paper is to find out the physical mechanism of FE phase induced by $V_O$.

# 2. OXYGEN VACANCIES IN THE RELAXORS AND THEIR CHARACTERISTIC FEATURES

Oxygen vacancies in $ABO_3$ ferroelectrics greatly impact their physical properties (see [6] and ref. therein) and the perovskite structure is able to conserve the structure stability even for high concentration of oxygen vacancies [7].

B cations are usually shifted from central position in the neighborhood of oxygen vacancy, because in $ABO_3$ structure the size of oxygen ions and so its vacancies used to be larger than cation ones. To compensate for the loss of oxygen negative charges the equivalent amount of $B^{4+}$ cations should be in a $B^{3+}$ state. The PZN-PLZT samples sintered in nitrogen atmosphere appeared to be black and opaque [6], because off-central $Ti^{4+}$ transforms into color center $Ti^{3+}$. Note, that $Ti^{3+}$ can create layers of ordered dipoles at large concentration of oxygen vacancies [8]. The electrons necessary for $Ti^{4+}$ into $Ti^{3+}$ transformation can be created from ionization of neutral oxygen vacancy $V_O \rightarrow V_O^{\bullet} + e$, $V_O^{\bullet} \rightarrow V_O^{\bullet\bullet} + e$, the $V_O^{\bullet}$ and $V_O^{\bullet\bullet}$ being positively charged vacancies. Uncharged vacancy $V_O$ represents dilatational center which creates local compressive strain.

Since generally the conductivity is mainly attributed to the electromigration of oxygen vacancies (see [9, 10]) in perovskite ferroelectrics, the measurements of *dc* conductivity temperature dependence of the NS and OA specimens were carried out in order to estimate of oxygen vacancies concentration and charge states. It was shown that NS specimens have several orders larger conductivity at high temperature (> 300 °C). Comparison of activation energies extracted from the conductivity temperature dependence had shown that concentration of oxygen vacancies in NS samples is high and attributed to $V_O^{\bullet\bullet}$ at low temperatures, while in OA samples the contribution of $V_O^{\bullet\bullet}$ and $V_O^{\bullet}$ were detected at high and low temperatures respectively. As to high temperature activation energy in NS specimens (0.95 eV), which appeared to be smaller than that in OA specimens (1.54 eV), the authors of [6] wrote, that this effect is attributed to higher concentration of oxygen vacancies in NS samples. To our mind this statement could be correct for the same type of vacancies in both samples e.g. compare $E_a$ for $V_O^{\bullet\bullet}$ in them. Singly ionized vacancy with $E_a = 0.95$ eV has also to be rejected because for such vacancy in AO sample $E_a = 0.31$ eV, that is much smaller than 0.95 eV in NS sample with much larger concentrations of oxygen vacancies. Because of this we supposed that uncharged vacancies can contribute as we will show later to activation energy 0.95 eV in NS samples, when complete compensation of loss oxygen negative charge 2e originates from two off-central $Ti^{3+}$ ions. Because this complex defect can be represented as $V_O^{\bullet\bullet} + 2Ti^{3+}$ it can be observed in high temperature region only. Therefore in NS sample with high concentration of oxygen vacancies we are faced with existence $V_O^{\bullet\bullet}$ and $V_O$.

Note that vacancies tend to accumulate in the vicinity of any inhomogeneities, surfaces and interfaces, since the energy of vacancies formation in such places can be much smaller than in the homogeneous volume [11, 12, 13, 14]. In the places of vacancies accumulation they can create sufficiently strong fields, which in turn can lead to new phases appearance in relaxors, for example, polar (ferroelectric) ones. On contrary, in the places where there are few vacancies the non-polar relaxor remains. So, the polar ferroelectric and nonpolar relaxor states coexistence can be realized in the case.

## 3. TEMPERATURE DEPENDENCE OF MODIFIED DIELECTRIC PERMITTIVITY IN PZN-PLZT NS SPECIMEN

It is well-known (see e.g. [15]) that the temperature dependence of dielectric permittivity of ferroelectric relaxor can be described by modified Curie-Weiss law:

$$\frac{1}{\varepsilon} - \frac{1}{\varepsilon_{max}} = \frac{1}{K}(T - T_M)^p. \qquad (1)$$

Here $p = 2$, $T_M$ marks the temperature of dielectric permittivity maximum, $K$ is a constant, while for normal ferroelectrics $T_M = T_C$, $p = 1$, $T_C$ is Curie temperature. This difference originates from the broken translational symmetry of the relaxors and modified Curie-Weiss law includes the frequency dependence in $\varepsilon$ and also in $T_M$ contrary to classical Curie-Weiss law for ferroelectrics. The pronounced frequency dependence in relaxors is known to be produced by broad relaxation time spectrum, described by Vogel-Fulcher law $1/\tau = (1/\tau_0)\exp(-U/(k(T-T_g)))$, where $T_g$ is freezing temperature. This and other peculiarities of relaxor ferroelectrics originate from random electric field, induced by substitutional disorder, presence of vacancies and other unavoidable defects (see e.g. [3] and ref. therein).

Deng et al [6] measured temperature dependence of PZN-PLZT relaxor dielectric permittivity for NS and AO samples and obtained respectively p = 1.53 and 1.91. The obtained data speaks in favor of the statement that the large concentration of oxygen vacancies induces ferroelectricity in NS sample, so that we are faced of its coexistence with relaxor state. This compound looks like $PMN_{1-x}PT_x$ solid solution, where one could observe so called morphotrophic region with relaxor and polar phases coexistance (see e.g. [16]). In what follows we will consider physical mechanism of induced ferroelectricity.

## 4. POSSIBLE MECHANISMS OF FERROELECTRICITY INDUCED BY OXYGEN VACANCIES

Let us consider briefly possible mechanisms of ferroelectricity appearance in NS samples of PZN-PLZT. As we discussed above the oxygen vacancies in this sample are uncharged $V_O$, singly and doubly positively charged $V_O^{\bullet}$ and $V_O^{\bullet\bullet}$ respectively. Because of necessity of loss oxygen negative charges compensation approximately equivalent amount of $Ti^{3+}$ off-central ions is another group of defects. Keeping in mind the electrostriction in disordered systems the elastic dipoles transform into electric ones.

An illustration of the oxygen vacancy related defect configurations in a tetragonal perovskite lattice structure is shown in **Fig. S1** in the Suppl. Mat. [17], adapted from Ref. [18]. The existence of electric dipoles will lead to appearance of ferroelectric phase due to indirect interaction of dipoles via soft optic mode [19], and the soft mode existence in the ferroelectric relaxors will be discussed later.

Allowing for that all the electric dipoles in the regions with sizes of order of correlation radius $r_c$ must be oriented, one can write the criterion of ferroelectric phase appearance as $Nr_c^3 \geq 1$, $N$ is concentration of dipoles. It was follows we will name the lowest concentration $N_c = r_c^{-3}$ correlation threshold.

Another possible mechanism of ferroelectricity in the relaxors can originate from inhomogeneous elastic field via flexoelectric effect, namely $P_i = f_{ijkl} \partial u_{kj}/\partial x_l$, where $P_i$ is electric polarization component, $\partial u_{ij}/\partial x_l$ is mechanical strain gradient, $f_{ijkl}$ is the tensor components of flexoelectric effect. Detailed consideration of this mechanism along with mechanical strain field originated from oxygen vacancies (Vegard mechanism) contributions to appearance of ferroelectricity in the relaxors will be performed in the next part.

## 5. VEGARD STRAINS CONTRIBUTION TO APPEARANCE OF FERROELECTRICITY IN RELAXORS

Gehring et al. [20] performed neutron inelastic scattering measurements of the lowest-energy transverse optic (TO) phonon branch in the relaxor Pb(Mg$_{1/3}$Nb$_{2/3}$)O$_3$ from 400 to 1100 K. Far above the Burns temperature $T_d$ = 620 K Gehring et al. observed well-defined propagating TO modes at all wave vectors $q$, and a zone center TO mode that softens in a manner consistent with that of a ferroelectric soft mode. Below $T_d$ the zone center TO mode is over-damped and its direct measurement becomes combersome. However Gehring et al. [21] supposed that this mode recovers as has been reported for PZN, where at 20 K TO mode was observed. The latter is very important for us because considered in [6] relaxor PZN-PLZT described by the formula 0.3Pb(Zn$_{1/3}$Nb$_{2/3}$)O$_3$–0.7(Pb$_{0.96}$La$_{0.04}$(Zr$_x$Ti$_{1-x}$)$_{0.99}$O$_3$) with the composition ($x$=0.52) near the morphotropic phase boundary, where La concentration is 4 % and so it has no relaxor properties, so that relaxor type behavior has to originate mainly from PZN. Therefore, we came to the conclusion about possibility to introduce hidden soft mode in the considered relaxor. This permits us to use LGD type free energy functional for quantitative consideration of ferroelectricity induced by oxygen vacancies in the relaxors. Note, that the same approach was used earlier in [22] for the description of the relaxor ferroelectric PLZT ceramics with 8 % and 9 % of La and PZN–4.5%PT single crystals.

Note, that applicability of standard Landau phenomenological model for calculation of electrocaloric effect in relaxor ferroelectrics in [23] is out of doubt because of induced by electric field polarization.

Gibbs potential density of relaxor ferroelectric materials having some hidden soft phonon polar mode [20, 21, 24] has the following form [25],

$$G = \frac{\alpha(T)}{2} P_i P_j + \frac{\alpha_{ijkl}}{2} P_i P_j P_k P_l + \frac{g_{ijkl}}{2} \frac{\partial P_i}{\partial x_j} \frac{\partial P_k}{\partial x_l} + \frac{F_{ijkl}}{2}\left(\sigma_{kl}\frac{\partial P_i}{\partial x_j} - P_i \frac{\partial \sigma_{kl}}{\partial x_j}\right),$$

$$- Q_{ijkl}\sigma_{ij}P_k P_l - \frac{s_{ijkl}}{2}\sigma_{ij}\sigma_{kl} + k_B T\; S(N_d, N_d^+) - P_i E_i - W_{ij} N_d(\vec{r})\sigma_{ij}$$

(2)

where $P_i$ are the components of polarization vector ($i$ =1, 2, 3) and $\sigma_{ij}$ is the elastic stress tensor. The summation is performed over all repeated indices. Dielectric stiffness coefficient $\alpha(T)$ is positive, because the intrinsic ferroelectricity is absent, but depends on temperature reflecting the fact that the hidden phonon mode could soften at negative absolute temperatures. This statement follows from abovementioned fact, that in PZN soft mode was observed at $T$ = 20 K, so that its frequency could be zero at negative temperature. Note, that extrapolation of PMN soft mode frequency to zero leads to $T_c \approx -150$ K. Matrix of the gradient coefficients $g_{ijkl}$ is positively defined. $Q_{ijkl}$ is the electrostriction tensor, $s_{ijkl}$ is the elastic compliances tensor, $F_{ijkl}$ is the forth-rank tensor of flexoelectric coupling. The configuration entropy function $S(x, y)$ is taken as $S(x, y) = y \ln(y/x) - y$ in the Boltzmann-Planck-Nernst approximation; $k_B$=1.3807×10$^{-23}$ J/K, where $T$ is the absolute temperature.

In Eq.(2), $E_i(\mathbf{r})$ denotes the internal electric field that satisfies electrostatic equation

$$\varepsilon_b \varepsilon_0 \frac{\partial E_i}{\partial x_i} + \frac{\partial P_i}{\partial x_i} = e(N_d^+ - n),$$

(3)

where $\varepsilon_b$ is background permittivity [26] and $\varepsilon_0$=8.85×10$^{-12}$ F/m is the universal dielectric constant, $e(N_d^+ - n)$ is the space charge density of singly ionized ionized vacancies and electrons, $e$=1.6×10$^{-19}$ C is an electron charge. The field $E_i(\mathbf{r})$ is induced by nonzero divergence of the bound charge $\frac{\partial P_i}{\partial x_i}$ (depolarization contribution) and space charge fluctuations related with of ionized vacancies (random field contribution).

Taking into account that the "net" random electric fields should be created by charged defects [4, 27], they are not the local fields originated around randomly distributed elastic dipoles due flexoelectric effect, but rather quenched by Imry-Ma scenario random fields [28]. It should be argued that the amount of charged vacancies it usually much smaller than the amount of uncharged vacancies, which are considered below as the main sources of ferroelectricity allowing for the Vegard mechanism. Actually, as a rule the concentration of the charged defects (ionized vacancies in our case) is of the order of several percents of the total defect concentration in the volume of the material, with the exception of near-surface charged layers, where their accumulation or depletion is possible [29].

The maximum number of vacancies observed in NS samples in Ref.[6] relates to uncharged oxygen vacancies (i.e. to elastic dipoles).

Equations of state $\partial G/\partial \sigma_{ij} = -u_{ij}$ determine the strains $u_{ij}$. Euler-Lagrange equations $\partial G/\partial P_i = 0$ determine the polarization components.

The last term in Eq.(2) is the **Vegard-type** concentration-deformation energy, $W_{ij} N_d \sigma_{ij}$, determined by the elastic defects (e.g., charged or electroneutral oxygen vacancies) with fluctuating concentration $N_d(\mathbf{r}) \cong \left\langle \sum_k \delta(\mathbf{r} - \mathbf{r}_k) \right\rangle \equiv \overline{N}_d + \delta N_d(\mathbf{r})$, where the equilibrium concentration is $\overline{N}_d$ and $\delta N_d(\mathbf{r})$ is the random variation. The variation $\delta N_d(\mathbf{r})$ is characterized by zero spatial average and nonzero mean square dispersion values, i.e. $\langle \delta N_d(\mathbf{r}) \rangle = 0$ and $\langle \delta N_d^2(\mathbf{r}) \rangle = n_d^2 > 0$. At that different cases $\overline{N}_d \gg n_d$, $\overline{N}_d \sim n_d$ and $\overline{N}_d \ll n_d$ are possible for oxygen vacancies. The average distance between defects centres 2R should be associated with the *average volume per inclusion* and so is defined from the relation $\frac{4\pi}{3} R^3 = \frac{1}{\overline{N}_d}$. The defect size $r_0$ is much smaller than the average distance R, e.g. $r_0$ is ionic radius $\sim (0.1 - 1)$Å$^3$ (see **Fig.1**).

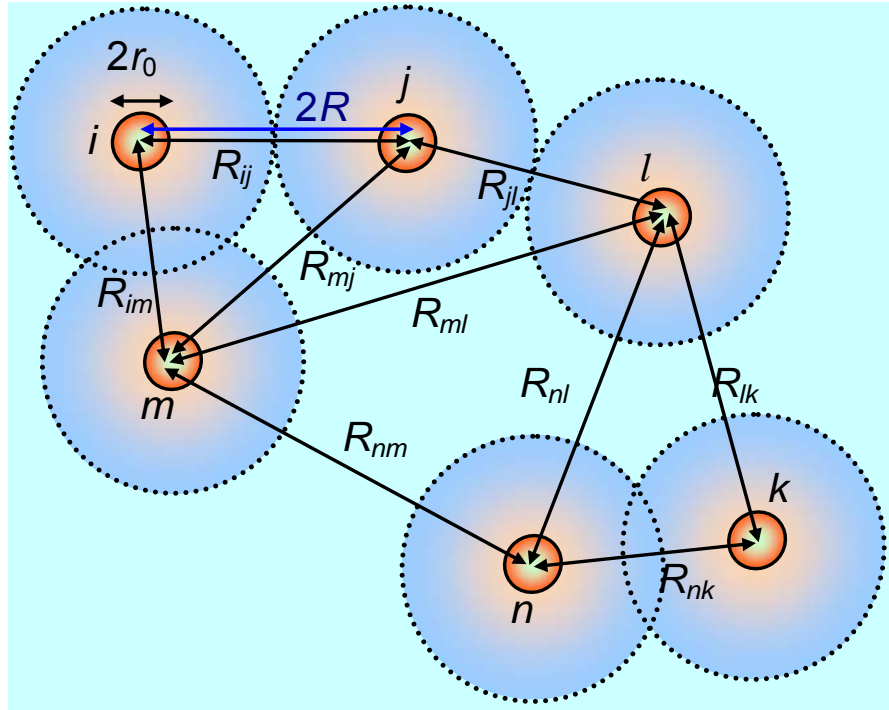

**FIGURE 1.** Schematics of spherical elastic defects with radius $r_0$ embedded into the matrix. The distance between defects "i" and "j" is $R_{ij}$. The average distance between defects is 2R. The average volume per one defect is $V = \frac{4\pi}{3} R^3$.

Nonzero components of the Vegard stresses induced by a spherically-symmetric elastic point defect (e.g. dilatation centre) located in the coordinate origin, $\mathbf{r} = 0$, in spherical coordinates have the form [30]

$$\sigma_{rr}^W(\mathbf{r}) = \frac{-W_{rr}}{2\pi(s_{11}-s_{12})r^3}, \qquad \sigma_{\theta\theta}^W(\mathbf{r}) = \sigma_{\varphi\varphi}^W(\mathbf{r}) = \frac{W_{\theta\theta}}{4\pi(s_{11}-s_{12})r^3} \qquad (4)$$

Equations (4) are derived under the assumption of isotropic and diagonal Vegard expansion tensor $W_{ij}$, $W_{ij} = W\delta_{ij}$. In general case the structure of Vegard expansion tensor $W_{ij}$ [31, 32, 33] (elastic dipole) is controlled by the symmetry (crystalline or Curie group symmetry) of the material. In Eqs.(4) the distance $r > r_0$. The elastic compliances tensor $s_{ij}$ is written in Voight notations.

Substitution of elastic fields (4) into the Eq.(2) leads to the renormalization of coefficient $\alpha(T) \to \alpha_{ij}^R$ by the joint action of electrostriction coupling and Vegard expansion,

$$\alpha_{kl}^R(T,\vec{r}) \approx \alpha(T) - 2Q_{ijkl}\sigma_{ij}^W(\mathbf{r}), \qquad (5)$$

One can see from Eq.(5) that the local polar state occurred under the condition $\alpha_{kl}^R < 0$ is not excluded in the spatial regions, where the defects concentration is high enough. Let us make some estimates.

Using ergodic hypothesis the averaging in Eq.(5) over the defects distribution function is reducing to the averaging over the defect partial volume, $V = \frac{4\pi}{3}R^3$, and gives the following expression:

$$\left\langle \alpha_{kl}^R(T,\overline{N}_d) \right\rangle \approx \alpha(T) - \frac{4}{3}Q_{ijkl}\widetilde{W}_{ij}\,\overline{N}_d \ln\!\left(\frac{3}{4\pi\overline{N}_d r_0^3}\right), \qquad (6a)$$

where $\widetilde{W}_{ij} = c_{ijkl}W_{kl}$ and $c_{ijkl}$ is the elastic stiffness tensor. Detailed derivation of Eq.(6) is listed in **Appendix B** of Ref.[17].

For a typical case $\alpha(T) = \alpha_T(T - T_C^*)$, where $T_C^*$ can be essentially smaller than room temperature or negative due to relaxor component, such as 30 % of PZN and disordering impurities 4 % of La. Pure PZT (52/48) has the Curie temperature of about 393°C. For a solid solution PZT (52/48) the coefficient $\alpha_T \approx 2.66 \times 10^5$ C$^{-2}$·m J/K, electrostriction coefficients $Q_{11}$=0.0966 m$^4$/C$^2$, $Q_{12}$= −0.0460 m$^4$/C$^2$, $Q_{44}$=0.08190 m$^4$/C$^2$, and elastic stiffness $c_{11}$=1.696×10$^{+11}$ Pa, $c_{12}$=0.819×10$^{+11}$ Pa [34, 35]. Vegard tensor is usually diagonal for oxygen vacancies in perovskites, but anisotropic, e.g., $W_{11} = 16.33$ Å$^3$ and $W_{22} = W_{33} = -8.05$ Å$^3$ for SrTiO$_3$ [33]. Note that the average concentration $\overline{N}_d$ should be much smaller than the value $2.25 \times 10^{28}$ m$^{-3}$ corresponding to one defect per unit cell with a size 4 Å. Thus, we obtain from expression (6a) that the renormalized transition temperature is equal to:

$$T_C^R \approx T_C^* + \frac{4}{3\alpha_T}\left[Q_{11}\widetilde{W}_{11} + Q_{12}\left(\widetilde{W}_{22} + \widetilde{W}_{33}\right)\right]\overline{N}_d \ln\left(\frac{3}{4\pi\overline{N}_d r_0^3}\right). \tag{6b}$$

Here $T_C^*$ is analog of Curie temperature for PZN relaxor, i.e. it originated from soft mode, observed at $T = 20$ K [21], and so $T_C^*$ has to be negative and can be estimated as $T_C^* \approx -(5 \div 100)$ K similarly to estimations made from soft mode observed in PMN [20]. Here we introduced the Vegard stress tensor components as follows $\widetilde{W}_{11} = c_{11}W_{11} + c_{12}(W_{22} + W_{33})$, $\widetilde{W}_{22} = c_{11}W_{22} + c_{12}(W_{11} + W_{33})$ and $\widetilde{W}_{33} = c_{11}W_{33} + c_{12}(W_{11} + W_{22})$.

It follows from Eq. (6b) that competition between contributions of the first and second terms can lead to $T_C^R > 0$ and so ferroelectricity could be observed. The shift of the Curie temperature $\Delta T_C$ vs. the average concentration of oxygen vacancies $\overline{N}_d$ calculated for several values of the Vegard tensor amplitude and defect sizes are shown **Fig. 2(a)** and **2(b)**, respectively. The shift monotonically increases with $\overline{N}_d$ increasing. Also it increases with Vegard coefficient increase and defect size decrease. One can see from **Figs. 2** that the polar phase is not excluded if the defects concentration $\overline{N}_d$ is high enough. Keeping in mind, that the oxygen vacancies concentration depends on technology [6], the special choice of $\overline{N}_d$ overcame also the correlation threshold necessary for the existence of normal switchable ferroelectric phase.

Noteworthy that the switchable ferroelectric polarization originates from the renormalization of Curie temperature given by Eq.(6b), and its two energetically equivalent (at zero external field) spontaneous values can be estimated as $P_{1,2} = \pm\sqrt{\frac{\alpha_T(T_C^R - T)}{\alpha_{11}}}$ in accordance with the mean-field approach. The values are the upper limit as estimated without gradient and depolarization effects, which decrease the polarization, and random fields, which can change it locally via disorder, as it will be analyzed in the next section. Note, that it was shown earlier that Vegard strains and stresses coupled with electrostriction and flexoelectricity indeed can induce a reversible ferroelectric polarization in nanosized ferroelectrics, which are paraelectric otherwise [36, 37].

It is important to underline that for intermediate concentration of oxygen vacancies the coexistence of relaxor state and ferroelectric phase can take place. It is not excluded that the result obtained by Deng et al [6] for NS samples with p=1.53 is characterizing neither ferroelectric nor relaxor and speaks in favor of the statement that we faced with the coexistence of different states, namely, of ordered ferroelectric and disordered relaxor.

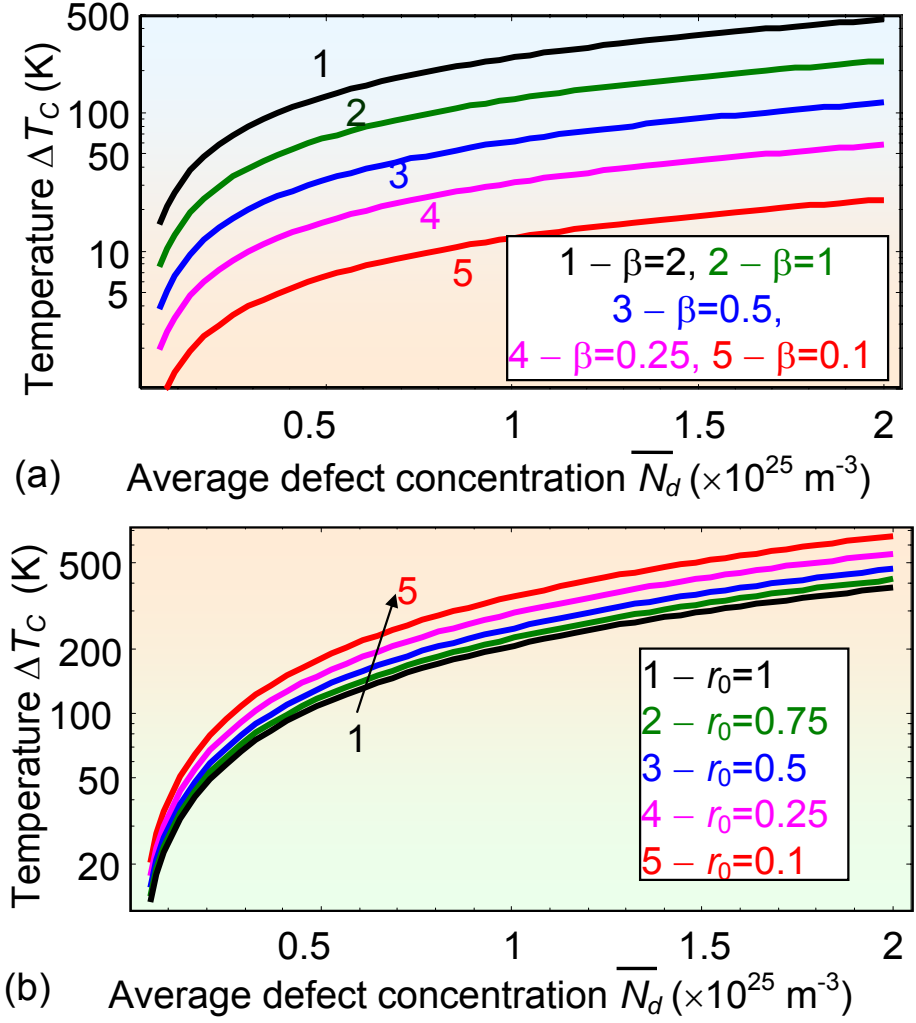

**FIGURE 2. (a)** The shift of the Curie temperature $\Delta T_C$ vs. the average concentration $\overline{N}_d$ of oxygen vacancies calculated for several values of the Vegard tensor amplitude β=2 (curve 1), β=0.5 (curve 2), β=0.5 (curve 3), β=0.25 (curve 4) and β=0.1 (curve 5). Vegard strain tensor $w_{ii} = \beta W_{ii}$, where $W_{11} = 16.33$ Å$^3$ and $W_{22} = W_{33} = -8.05$ Å$^3$. Defect size is $r_0 = 0.5$ Å$^3$. **(b)** The shift of the Curie temperature $\Delta T_C$ vs. the average concentration of oxygen vacancies calculated for several values of the defect size $r_0 = 1$ Å$^3$ (curve 1), $r_0 = 0.75$ Å$^3$ (curve 2), $r_0 = 0.5$ Å$^3$ (curve 3), $r_0 = 0.25$ Å$^3$ (curve 4) and $r_0 = 0.1$ Å$^3$ (curve 5). Vegard strain tensor components are $W_{11} = 16.33$ Å$^3$ and $W_{22} = W_{33} = -8.05$ Å$^3$.

## 6. LOCAL POLARIZATION AND ELECTRIC FIELD INDUCED BY FLEXO-CHEMICAL COUPLING IN A RELAXOR

Let us estimate the polarization and electric fields variations induced by the joint action of Vegard stresses and flexoelectric coupling. Equations of state $\partial G/\partial \sigma_{ij} = -u_{ij}$ give the strains $u_{ij}$ as:

$$u_{ij} = s_{ijkl}\sigma_{kl} + W_{ij}\delta N_d - F_{ijkl}\frac{\partial P_l}{\partial x_k} + Q_{ijkl}P_k P_l \ . \tag{7a}$$

Since the static equation of mechanical equilibrium, $\partial \sigma_{ij}/\partial x_j = 0$, should be valid, Eq.(7a) transforms into Lame-type equation for elastic displacement $U_i$:

$$c_{ijkl}\frac{\partial^2 U_l}{\partial x_j \partial x_k} = \frac{\partial}{\partial x_j}\left(\sigma_{ij}^W - f_{ijkl}\frac{\partial P_l}{\partial x_k} + q_{ijkl}P_k P_l\right), \quad (7b)$$

where $c_{ijkl}$ are elastic stiffness, Vegard stress $\sigma_{mn}^W = c_{ijmn}W_{ij}\delta N_d$, $q_{mnkl} = c_{ijmn}Q_{ijkl}$ is electrostriction stress tensor and $f_{mnkl} = c_{ijmn}F_{ijkl}$ is the flexocoupling stress tensor.

Minimization of the Gibbs potential (2) with respect to $P_j$ leads to the Landau-Ginzburg-Devonshire type equations for ferroelectric polarization components:

$$(\alpha\delta_{ij} - 2\sigma_{mn}Q_{mnij})P_j + \alpha_{ijkl}P_j P_k P_l - g_{ijkl}\frac{\partial^2 P_k}{\partial x_j \partial x_l} = F_{mnli}\frac{\partial \sigma_{mn}}{\partial x_l} + E_i \quad (8)$$

The electric field $E_i$ is the sum of internal (depolarizing and spatially random) and probing external fields, $E_i = \delta E_i + E_i^{ext}$, which should be found self-consistently from Eq.(3). Further we put $E_i^{ext} = 0$ being interested in the random field only. For the case Eq.(3) can be rewritten in the form of Poisson equation for electric potential $\varphi$:

$$\varepsilon_b \varepsilon_0 \frac{\partial^2 \varphi}{\partial x_i^2} = \frac{\partial P_i}{\partial x_i} + e(\delta N_d^+ - \delta n), \quad (9)$$

The conventional relation $E_i = -\partial \varphi / \partial x_i$ is valid. The isotropic relative permittivity $\varepsilon_b$ that can be high enough $\sim(10^2 - 10^3)$ for relaxor ferroelectrics, while for normal ferroelectrics it is a of background that is not more than 10. The total electroneutrality condition $\overline{N}_d^+ = \overline{n}$ (valid at $E_i^{ext} = 0$) is already used in Eq.(9).

In **Appendix A** of Ref.[17] we solved the linearized system of Eqs.(7b), (8) and (9) using the method outlined in Ref.[38]. The Fourier **k**-spectrum of the internal electric field variation (depolarizing by nature and random as induced by random variation of defect concentration) was found in Debye approximation from Eq.(9). It has the form

$$\delta\widetilde{E}_j(\mathbf{k}) \approx \frac{-k_m k_j \delta\widetilde{P}_m}{\varepsilon_b\varepsilon_0(k^2 + R_d^{-2})}, \quad (10a)$$

where $R_d$ is the screening radius that depends on temperature and average concentration of defects as $R_d = \sqrt{\dfrac{\varepsilon_b\varepsilon_0 k_B T}{2e^2 \overline{N}_d}}$. The Fourier **k**-spectrum of the polarization variation induced by the randomly distributed vacancies due to Vegard stresses and flexoelectric coupling has the following form:

$$\delta\widetilde{P}_j(\mathbf{k}) \approx if_{mnli}k_l k_n k_{j'}S_{mi'}(\mathbf{k})\widetilde{\chi}_{ij}(\mathbf{k})\widetilde{\sigma}_{i'j'}^W, \quad (10b)$$

Random stresses are related with the random vacancies sites as $\sigma_{mn}^{W} = c_{ijmn} W_{ij} \delta N_d$. The converse tensors of dielectric susceptibility $\tilde{\chi}_{ij}^{-1}(\mathbf{k})$ and elastic matrix $S_{ik}^{-1}(\mathbf{k})$ have the form:

$$\tilde{\chi}_{ij}^{-1}(\mathbf{k}) \approx \alpha \delta_{ij} + g_{ipjl} k_p k_l + \frac{k_i k_j}{\varepsilon_b \varepsilon_0 (k^2 + R_d^{-2})}, \qquad S_{ik}^{-1}(\mathbf{k}) = c_{ijkl} k_l k_j, \qquad (10c)$$

where $c_{ijkl}$ are elastic stiffness. The second term $g_{ipjl} k_p k_l$ in expression for $\tilde{\chi}_{ij}^{-1}(\mathbf{k})$ originates from the polarization gradient, and the third term $\frac{k_i k_j}{\varepsilon_b \varepsilon_0 (k^2 + R_d^{-2})}$ originates from depolarization effects calculated in Debye approximation. Also it was shown earlier (see e.g. Ref.[3] and refs [14]-[16] therein), that the contribution of the gradient term is important for nanosized structures, while for homogeneous macro-sized structures this contribution is negligible.

The variations of random electric field (10a) and local polarization (10b) are not related with the ferroelectricity induced by uncharged vacancies with the average concentration $\overline{N}_d$ considered in the mean-field approach in section 5. The ferroelectric polarization is proportional to nonzero average concentration of vacancies $\overline{N}_d$. The variation $\delta \vec{P}(\mathbf{r})$ is, of course, not switchable and its average value is zero $\langle \delta \vec{P}(\mathbf{r}) \rangle = 0$, since $\langle \delta N_d(\mathbf{r}) \rangle = 0$.

That is why it makes sense to calculate standard (mean square) deviations of local polarization $\langle \delta \vec{P}^2(\mathbf{r}) \rangle$ and electric field $\langle \delta \vec{E}^2(\mathbf{r}) \rangle$, which are proportional to the dispersion of defect concentration $\langle \delta N_d^2(\mathbf{r}) \rangle$. Note that analytical estimates of the values $\sqrt{\langle \delta \vec{P}^2(\mathbf{r}) \rangle}$ and $\sqrt{\langle \delta \vec{E}^2(\mathbf{r}) \rangle}$ are possible only for the case of simplest spherically symmetric dilatation centre in an isotropic surrounding, while more realistically it is cubic. Analytical expressions for $\sqrt{\langle \delta \vec{P}^2(\mathbf{r}) \rangle}$ and $\sqrt{\langle \delta \vec{E}^2(\mathbf{r}) \rangle}$ are derived in **Appendix A** of Ref.[17], they are:

$$\sqrt{\langle \delta \vec{P}^2(\mathbf{r}) \rangle} \cong \frac{|f_{11} W_{eff} k_c| \cdot \sqrt{\langle \delta N_d^2 \rangle}}{\left| \alpha_T (T - T_C^*) + g_{11} k_c^2 + \frac{k_c^2 R_d^2}{\varepsilon_b \varepsilon_0 (1 + k_c^2 R_d^2)} \right|}, \qquad (11a)$$

$$\sqrt{\langle \delta \vec{E}^2(\mathbf{r}) \rangle} \cong \frac{k_c^2 R_d^2 |f_{11} W_{eff} k_c| \sqrt{\langle \delta N_d^2 \rangle}}{\varepsilon_b \varepsilon_0 (1 + k_c^2 R_d^2) \left| \alpha_T (T - T_C^*) + g_{11} k_c^2 + \frac{k_c^2 R_d^2}{\varepsilon_b \varepsilon_0 (1 + k_c^2 R_d^2)} \right|}, \qquad (11b)$$

where the effective Vegard coefficient $W^{eff} = W_{11}\left(1 + 2\frac{c_{12}}{c_{11}}\right)$ and wave vector $k_c \cong \xi\sqrt[3]{1/\overline{N}_d}$ characteristic for long-range correlations are introduced. Since $k_c$ defines the period of long-range correlations the dimensionless parameter $\xi$ should in order of unity and $k_c \ll \frac{1}{r_0}$ as anticipated.

Let us perform numerical estimates of the gradient term ($g_{11}k_c^2$) and depolarization ($\frac{k_c^2 R_d^2}{\varepsilon_b \varepsilon_0 (1 + k_c^2 R_d^2)}$) contributions in Eqs.(11) for typical values of parameters: the gradient coefficient $g_{11} = (0.1 - 5) \times 10^{-10}$ m$^3$/F, inverse Curie-Weiss constant $\alpha_T \approx 2.66 \times 10^5$ C$^{-2}$·m J/K, virtual Curie temperature $T_C^* \approx -(5 \div 100)$ K and relative permittivity $\varepsilon_b \sim 10^2$ for relaxor ferroelectrics. For chosen parameters "dressed" Debye screening radius $R_d$ can be smaller than 0.8 nm at 293 K at $\overline{N}_d = 10^{25}$ m$^{-3}$, and $\xi \sim 1$. For these values the gradient and depolarization contributions are, $g_{11}k_c^2 \sim \times 10^5$ C$^{-2}$·m J and $\frac{k_c^2 R_d^2}{\varepsilon_b \varepsilon_0 (1 + k_c^2 R_d^2)} \sim 0.8 \times 10^7$ C$^{-2}$·m J, respectively. These values are significantly smaller than the soft mode contribution $|\alpha_T(T - T_C^*)| \sim (0.81 - 1.06) \times 10^8$ C$^{-2}$·m J at room temperature.

Since it is not excluded that for other parameters (e.g. for smaller $\varepsilon_b$ and $\overline{N}_d$) the depolarization contribution can be compatible and even higher than the soft mode one, we note that Eqs.(10)-(11) are applicable for all parameters.

The dependences of mean squire deviation of local polarization $\sqrt{\langle \delta \vec{P}^2(\mathbf{r}) \rangle}$ and electric field $\sqrt{\langle \delta \vec{E}^2(\mathbf{r}) \rangle}$ on the dispersion $\sqrt{\langle \delta N_d^2 \rangle}$ of the vacancies concentration fluctuations are shown in **Figs.3(a)** and **3(b)** for several values of dimensionless parameter $\xi$ within the range (0.1 – 10). The values $\sqrt{\langle \delta \vec{P}^2(\mathbf{r}) \rangle}$ and $\sqrt{\langle \delta \vec{E}^2(\mathbf{r}) \rangle}$ monotonically increase with $\sqrt{\langle \delta N_d^2 \rangle}$ increase.

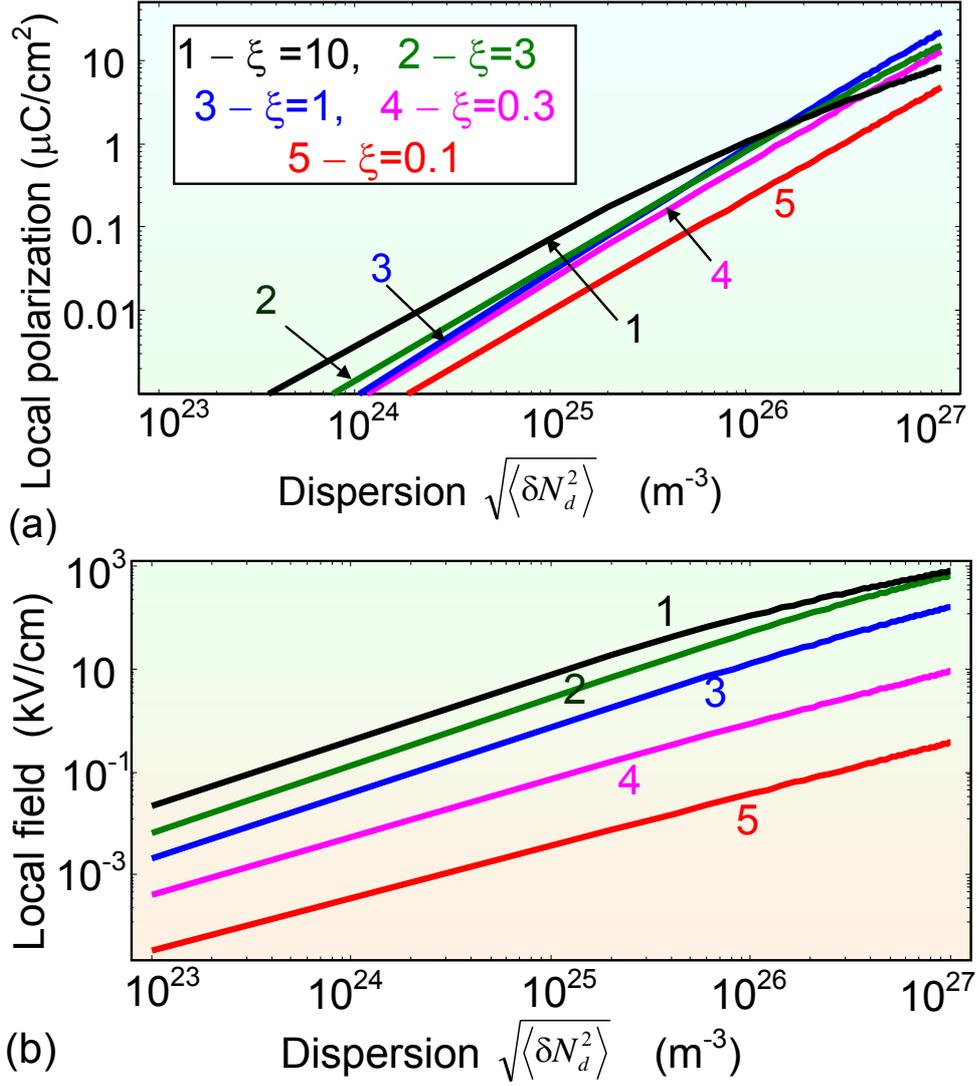

**FIGURE 3.** Mean squire deviations of **(a)** local polarization $\sqrt{\langle \delta \vec{P}^2(\mathbf{r}) \rangle}$ and **(b)** electric field $\sqrt{\langle \delta \vec{E}^2(\mathbf{r}) \rangle}$ vs. the dispersion $\sqrt{\langle \delta N_d^2 \rangle}$ of vacancies concentration fluctuations are shown in log-log scale for several values of parameter $\xi = 10$ (curve 1), $\xi = 3$ (curve 2), $\xi = 1$ (curve 3), $\xi = 0.3$ (curve 4) and $\xi = 0.1$ (curve 5). Vegard strain tensor strength $W = 10$ Å$^3$ and flexocoupling constant $f_{11} = 4$ V estimated from Kogan model [39] and temperature T=300 K.

To resume the section, the dependence of mean square deviation of polarization $\langle \delta P^2(\mathbf{r}) \rangle$ and internal electric field $\langle \delta E^2(\mathbf{r}) \rangle$ on concentration of oxygen vacancies $\sqrt{\langle \delta N_d^2 \rangle}$ [shown in **Figs. 3**] demonstrated that flexo-chemical coupling essentially contributes to local polarization and internal electric field. This speaks in favor of statement about existence of dipole glass state. So that the proposed model explain and quantify some features of the dipole glass and relaxor states coexistence [19], that has been observed experimentally [40, 41, 42].

# 7. DISCUSSION AND CONCLUSIONS

The main experimental fact obtained in [6] originated from measurements of temperature dependence of PZN-PLZT relaxor dielectric permittivity for NS and AO samples. Authors obtained respectively $p$ = 1.53 and 1.91 in Eq. (1). These data speaks in favour of statement that in NS samples large concentration of oxygen vacancies induced ferroelectricity, so that we are faced with its coexistence with relaxor state of PZN. The obtained result resembles $PMN_{1-x}PT_x$ compound for which $p$ = 1.53 can be obtained for $x$ = 0.5 approximately. For this concentration transition to ferroelectric phase takes place at $T_C$ =250°C > 0 in PMN-PT (see e.g. [43]). The main task of our consideration was to find out the physical mechanism, that can be in response of oxygen vacancies induced ferroelectricity in PZN-PLZT relaxor. As a matter of fact $(Pb, La)Zr_{0.52}Ti_{0.48}O_3$ with La content below 10 % is known to be in ferroelectric phase [44], that could suppress relaxor disorder of PZN. However allowing for the value of parameter p = 1.91 for OA samples annealed in oxygen is very close to p = 2, that is characteristic value for relaxors, we neglected PLZT contribution for NS samples, where p = 1.53 and the concentration of oxygen vacancies is large (see Introduction). Keeping in mind that oxygen vacancies are elastic dipoles, which influence used to be considered as Vegard mechanism, we performed the calculations with the help of defect concentration distribution function and obtained Eq.(6). The results depicted in **Figs. 2** had shown, that at some average concentration of oxygen vacancies $\overline{N}_d$ their contribution can be larger than negative value of temperature $T_C^*$ of a relaxor and so we obtained positive transition temperature characteristic for a ferroelectric. For transformation of the relaxors into ferroelectrics one need large enough concentration of oxygen vacancies and Vegard tensor amplitude. Unfortunately exact value of negative Curie temperature $T_C^*$ is not known and we have to discuss some estimations only. It is obvious that even large enough value of $T_C^*$ can be overcome by special choice of oxygen vacancies concentrations and other parameters. The latter is very important because the vacancies concentration has to be larger than the correlation threshold necessary for the existence of normal ferroelectric reversible phase. In such a case at $T < T_C^R$ (ferroelectric phase) all the properties could be calculated by conventional way on the base of free energy $G = \dfrac{\alpha_T(T_C^R - T)}{2}P^2 + \dfrac{\alpha_{11}}{4}P^4$, so that e.g. polarization $P^2 = \dfrac{\alpha_T(T_C^R - T)}{\alpha_{11}}$, while at $T > T_C^R$ the polarization is zero, $P = 0$, and we again have the relaxor. For this case in p. 6 we consider local polarization and electric field induced by Vegard and flexoelectric effects (flexo-chemical coupling). The dependence of mean square deviation of polarization $\langle \delta P^2(\mathbf{r}) \rangle$ and internal electric field $\langle \delta E^2(\mathbf{r}) \rangle$ on concentration of oxygen vacancies $\sqrt{\langle \delta N_d^2 \rangle}$ [shown in **Figs. 3**] speaks in favor of statement about existence of dipole glass state. So that the proposed model can explain and quantify some features of the dipole glass and relaxor states coexistence, that has been observed experimentally [19, 40, 41, 42].

Keeping in mind the accumulation of oxygen vacancies in vicinity of different inhomogeneity we came to the conclusion about oxygen vacancies concentration inhomogeneity. In such a case one can expect coexistence of relaxor state and ferroelectricity. It is not excluded that this interesting phenomena was observed by Deng [6] for NS samples with p=1.53. It lays approximately at the same distance from p=2 for relaxors and p=1 for ferroelectrics. Therefore, at some concentration of oxygen vacancies and T<$T_C^R$ we are faced with morphotrophic region in PZN-PLZT.

To resume, we have shown that the transition to a ferroelectric phase can be induced in a relaxor by the influence of oxygen vacancies being elastic dipoles due to the joint action of electrostrictive and Vegard couplings at some large enough concentration of the vacancies. In the regions where the concentration of vacancies is low, the local polarization and electric field could be induced by the flexo-chemical coupling in dependence on the concentration of oxygen vacancies. Because of inhomogeneity of vacancies concentration the coexistence of ferroelectricity and relaxor state can be expected.

**Acknowledgements.** The authors express their deep gratitude to the Referees for very useful suggestions and comments.

**Authors contribution.** M.D.G. generated the research idea, stated the problem and wrote section 1-4, discussion and conclusions of the manuscript. A.N.M. wrote the sections 5-6 and Appendix A with corresponding calculations and, jointly with E.A.E., generated figures. E.A.E., G.L. and S.V.K. worked on the results discussion and manuscript improvement.

# REFERENCES

[1] Leslie Eric Cross, "Ferroelectric materials for electromechanical transducer applications." Materials chemistry and physics 43, no. 2: 108-115 (1996).

[2] Seung-Eek Park, and Thomas R. Shrout. "Characteristics of relaxor-based piezoelectric single crystals for ultrasonic transducers." IEEE Transactions on Ultrasonics, Ferroelectrics, and Frequency Control 44, no. 5: 1140-1147 (1997).

[3] M.D. Glinchuk, A.V. Ragulya, V.A. Stephanovich. Nanoferroics. Springer Series in Materials Science. Dordrecht: Springer (2013), 378 p. (ISBN 978-94-007-5992-3)

[4] M. D. Glinchuk, and R. Farhi. "A random field theory based model for ferroelectric relaxors." Journal of Physics: Condensed Matter **8** (37), 6985 (1996).

[5] M.D.Glinchuk, Relaxor ferroelectrics: from Cross superparaelectric model to random field theory, British Ceramic Transactions, **103**, N2, 76-82 (2004).

[6] G. Deng, G. Li, A. Ding, Q. Yin, "Evidence for oxygen vacancy inducing spontaneous normal-relaxor transition in complex perovskite ferroelectrics" *Applied physics letters* **87**, 192905 (2005).

[7] S. Steinsvik, R. Bugge, J. Gjonnes, J. Tafto, and T. Norby, "The defect structure of $SrTi_{1-x}Fe_xO_{3-y}$ (x= 0–0.8) investigated by Electrical Conductivity Measurements and Electron Energy Loss Spectroscopy (EELS)" J. Phys. Chem. Solids 58, 969 (1997).

[8] D. Woodward, Ian M. Reaney, Gaiying Y. Yang, Elizabeth C. Dickey, and Clive A. Randall. "Vacancy ordering in reduced barium titanate." *Applied physics letters* 84, no. 23: 4650-4652 (2004)..

[9] Nava Setter and L. E. Cross, " The role of B-site cation disorder in diffuse phase transition behavior of perovskite ferroelectrics " J. Appl. Phys. **51**, 4356 (1980).

[10] C. Ang, Z. Yu, and L. E. Cross, "Oxygen-vacancy-related low-frequency dielectric relaxation and electrical conduction in Bi: SrTiO 3." Phys. Rev. B 62, 228 (2000).

[11] Fenggong Wang, Zhiyong Pang, Liang Lin, Shaojie Fang, Ying Dai, and Shenghao Han. "Magnetism in undoped MgO studied by density functional theory." Phys. Rev. B 80 144424 (2009).

12 J. Carrasco, F. Illas, N. Lopez, E. A. Kotomin, Yu. F. Zhukovskii, R. A. Evarestov, Yu. A. Mastrikov, S. Piskunov, and J. Maier. "First-principles calculations of the atomic and electronic structure of F centers in the bulk and on the (001) surface of Sr Ti O 3." Phys. Rev. B 73, 064106 (2006).

[13] H. Jin, Y. Dai, BaiBiao Huang, and M.-H. Whangbo, "Ferromagnetism of undoped GaN mediated by through-bond spin polarization between nitrogen dangling bonds." Appl. Phys. Lett. 94, 162505 (2009).

[14] In particular, density functional calculations show that the energy of vacancy formation on the surface is lower than in the bulk on about 3 eV for GaN [13] and 0.28 eV for MgO [11].

[15] C. Ziebert, H. Schmitt, J. K. Krüger, A. Sternberg, and K. H. Ehses, "Grain-size-induced relaxor properties in nanocrystalline perovskite films." Phys. Rev. B 69, 214106 (2004).

[16] Hu Cao, Feiming Bai, Jiefang Li, D. Viehland, Guangyong Xu, H. Hiraka, and G. Shirane. "Structural phase transformation and phase boundary/stability studies of field-cooled Pb ( Mg 1 ∕ 3 Nb 2 ∕ 3 O 3 ) – 32 % Pb Ti O 3 crystals" Journal of Applied Physics 97, 094101 (2005).

[17] Supplementary materials with Solution of linearized system of Euler-Lagrange equations. Calculations of correlators. [URL will be provided by Publisher]

[18] Erhart, Paul, and Karsten Albe. "Dopants and dopant–vacancy complexes in tetragonal lead titanate: A systematic first principles study." *Computational Materials Science* **103** (2015): 224-230

[19] B. E. Vugmeister, M. D. Glinchuk, Dipole glass and ferroelectricity in random-site electric dipole systems, Rev. Mod. Phys., **62**, 993-1026 (1990).

[20] P. M. Gehring, S. Wakimoto, Z-G. Ye, and G. Shirane. "Soft mode dynamics above and below the Burns temperature in the relaxor Pb (Mg 1/3 Nb 2/3) O 3." Physical review letters 87, 277601 (2001).

[21] P. M. Gehring, S.-E. Park, and G. Shirane, "Dynamical effects of the nanometer-sized polarized domains in Pb (Zn 1/3 Nb 2/3) O 3." Phys. Rev. B 63, 224109 (2001).

[22] Andrei Kholkin, Anna Morozovska, Dmitry Kiselev Igor Bdikin, Brian Rodriguez Pingping Wu, Alexei Bokov, Zuo-Guang Ye, Brahim Dkhil, Long-Qing Chen, Marija Kosec Sergei V. Kalinin "Surface Domain Structures and Mesoscopic Phase Transition in Relaxor Ferroelectrics" Advanced Functional Materials 21, No 11, 1977–1987 (2011).

[23] R. Pirc, Z. Kutnjak, R. Blinc, and Q. M. Zhang. " Electrocaloric effect in relaxor ferroelectrics " Journal of Applied Physics, 110, , 074113 (2011).

[24] B.J. Rodriguez, S. Jesse, A.N. Morozovska, S.V. Svechnikov, D.A. Kiselev, A.L. Kholkin, A.A. Bokov, Z.-G. Ye, S.V. Kalinin, Real space mapping of polarization dynamics and hysteresis loop formation in relaxor-ferroelectric PbMg1/3Nb2/3O3–PbTiO3 solid solutions. J. Appl. Phys. 108, 042006 (2010).

[25] Anna N. Morozovska, Eugene A. Eliseev, Yuri A. Genenko, Ivan S. Vorotiahin, Maxim V. Silibin, Ye Cao, Yunseok Kim, Maya D. Glinchuk, and Sergei V. Kalinin. Flexocoupling impact on the size effects of piezo-response and conductance in mixed-type ferroelectrics-semiconductors under applied pressure. *Phys. Rev.* B 94, 174101 (2016)

[26] A.K. Tagantsev, and G. Gerra, "Interface-induced phenomena in polarization response of ferroelectric thin films." J. Appl. Phys. **100**, 051607 (2006)

[27] V. Westphal, W. Kleemann, and M. D. Glinchuk. "Diffuse phase transitions and random-field-induced domain states of the ''relaxor'' ferroelectric PbMg$_{1/3}$Nb$_{2/3}$O$_3$." Physical Review Letters **68**, no.6: 8476 (1992).

[28] Yoseph Imry and Shang-keng Ma. Random-Field Instability of the Ordered State of Continuous Symmetry. Phys. Rev. Lett. **35**, 1399 (1975)

[29] B. I. Shklovsky and A. L. Efros. "Electronic Characteristics of Doping Semiconductors." *Science, Moscow* (1979): 216

[30] Cristian Teodosiu, "Elastic Models of Crystal Defects" (Berlin, Springer-Verlag 1982).

[31] G. Catalan and James F. Scott. "Physics and applications of bismuth ferrite." Adv. Mater. **21**, 2463-2485 (2009).

[32] X. Zhang, A. M. Sastry, W. Shyy, "Intercalation-induced stress and heat generation within single lithium-ion battery cathode particles." J. Electrochem. Soc. **155**, A542 (2008).

[33] Daniel A. Freedman, D. Roundy, and T. A. Arias, "Elastic effects of vacancies in strontium titanate: Short- and long-range strain fields, elastic dipole tensors, and chemical strain." Phys. Rev. B **80**, 064108 (2009).

[34] M.J. Haun, E. Furman, S.J. Jang, H.A. McKinstry, and L. E. Cross. "Thermodynamic theory of PbTiO3". J. Appl. Phys. **62**, 3331 (1987).

[35] K. Rabe, Ch.H. Ahn, J.-M. Triscone (Eds.) *Physics of Ferroelectrics: A Modern Perspective*, Springer 2007 (L.-Q. Chen, Appendix A. Landau Free-Energy Coefficients).

[36] A.N. Morozovska, I.S. Golovina, S.V. Lemishko, A.A. Andriiko, S.A. Khainakov, and E.A. Eliseev. Effect of Vegard strains on the extrinsic size effects in ferroelectric nanoparticles Physical Review B 90, 214103 (2014)

[37] Anna N. Morozovska and Maya D. Glinchuk. Reentrant phase in nanoferroics induced by the flexoelectric and Vegard effects. *J. Appl. Phys.* 119, 094109 (2016)

[38] Anna N. Morozovska, Yulian M. Vysochanskii, Olexandr V. Varenik, Maxim V. Silibin, Sergei V. Kalinin, and Eugene A. Eliseev. Flexocoupling impact on the generalized susceptibility and soft phonon modes in the ordered phase of ferroics. Physical Review B 92 (9), 094308 (2015). http://arxiv.org/abs/1507.01108

[39] Sh. M. Kogan, Piezoelectric effect during inhomogeneous deformation and acoustic scattering of carriers in crystals Sov. Phys. Solid State 5, 2069 (1964).

[40] Shvartsman, V. V., A. L. Kholkin, A. Orlova, D. Kiselev, A. A. Bogomolov, and A. Sternberg. "Polar nanodomains and local ferroelectric phenomena in relaxor lead lanthanum zirconate titanate ceramics." *Applied Physics Letters* 86, no. 20 (2005): 202907.

[41] Shvartsman, V. V., J. Dec, T. Łukasiewicz, A. L. Kholkin, and W. Kleemann. "Evolution of the polar structure in relaxor ferroelectrics close to the Curie temperature studied by piezoresponse force microscopy." *Ferroelectrics* 373, no. 1 (2008): 77-85.

[42] Kleemann, W., V. V. Shvartsman, P. Borisov, and A. Kania. "Coexistence of antiferromagnetic and spin cluster glass order in the magnetoelectric relaxor multiferroic PbFe 0.5 Nb 0.5 O 3." *Physical review letters* 105, no. 25 (2010): 257202.

[43] Thomas R. Shrout, Zung P. Chang, Namchul Kim, and Steven Markgraf. "Dielectric behavior of single crystals near the (1− X) Pb (Mg1/3Nb2/3) O3-(x) PbTiO3 morphotropic phase boundary." Ferroelectrics Letters Section 12, no. 3: 63-69 (1990).

[44] Gene H. Haertling. "PLZT electrooptic materials and applications—a review" Ferroelectrics, 75, 25-55 (1987).

## SUPPLEMENTAL MATERIAL

## APPENDIX A. Linearized solution of Euler-Lagrange equations for fluctuations of electric polarization, field and strain. Calculations of correlations.

Expression for the Landau-Ginzburg-Devonshire type Gibbs potential has the following form:

$$G = \int_V d^3r \left( \begin{array}{c} \dfrac{\alpha(T)}{2} P_i P_j + \dfrac{\alpha_{ijkl}}{2} P_i P_j P_k P_l + \dfrac{g_{ijkl}}{2} \dfrac{\partial P_i}{\partial x_j} \dfrac{\partial P_k}{\partial x_l} + \dfrac{F_{ijkl}}{2} \left( \sigma_{kl} \dfrac{\partial P_i}{\partial x_j} - P_i \dfrac{\partial \sigma_{kl}}{\partial x_j} \right) \\ - Q_{ijkl} \sigma_{ij} P_k P_l - \dfrac{s_{ijkl}}{2} \sigma_{ij} \sigma_{kl} - P_i E_i - W_{ij} \delta N_d \sigma_{ij} + k_B T \, S(N_d, N_d^+) \end{array} \right) \quad (A.1)$$

Hereinafter summation is performed over all repeating indexes; $P_i$ is electric polarization. The expansion coefficient $\alpha$ is temperature dependent as $\alpha = \alpha_T (T - T_C)$, where $T$ is the absolute temperature, $T_C < 0$ since the intrinsic ferroelectricity is absent. Elastic stress tensor is $\sigma_{mn}$, $Q_{mnij}$ is the electrostriction strain tensor, $F_{mnli}$ is the flexoelectric effect tensor. The higher order coefficients $\alpha_{ijkl}$ are regarded temperature independent; $g_{ijkl}$ are gradient coefficients tensor, $s_{ijkl}$ are elastic compliances. Also we introduce the fluctuations of the Vegard strain, $W_{ij} \delta N_d$, which are proportional to the fluctuations of vacancies concentration $\delta N_d(\mathbf{r}) = N_d(\mathbf{r}) - \overline{N}_d$ from the average value and Vegard expansion tensor is $W_{ij}$ (that is typically isotropic and diagonal, $W_{ij} = W \delta_{ij}$, in disordered ferroelectrics)

Minimization of Eq.(A.1) with respect to $P_i$ leads to the static equation of state for ferroelectric polarization:

$$(\alpha \delta_{ij} - 2 \sigma_{mn} Q_{mnij}) P_j + \alpha_{ijkl} P_j P_k P_l - g_{ijkl} \dfrac{\partial^2 P_k}{\partial x_j \partial x_l} = F_{mnli} \dfrac{\partial \sigma_{mn}}{\partial x_l} + E_i \quad (A.2)$$

The quasi-static lectric field $E_i \equiv -\partial \varphi / \partial x_i$ $E_i$ can be found self-consistently from the electric potential $\varphi$ that satisfies Poisson equation,

$$\varepsilon_b \varepsilon_0 \dfrac{\partial^2 \varphi}{\partial x_i^2} = \dfrac{\partial P_i}{\partial x_i} - e(\delta N_d^+ - \delta n), \quad (A.3)$$

where $\varepsilon_b$ is background permittivity and $\varepsilon_0 = 8.85 \times 10^{-12}$ F/m is the universal dielectric constant, $e(\delta N_d^+ - \delta n)$ is the space charge density of ionized vacancies and electrons, $e = 1.6 \times 10^{-19}$ C is the electron charge.

Equations of state $\partial G / \partial \sigma_{ij} = -u_{ij}$ determine the strains $u_{ij}$:

$$u_{ij} = s_{ijkl} \sigma_{kl} + W_{ij} \delta N_d - F_{ijkl} \dfrac{\partial P_l}{\partial x_k} + Q_{ijkl} P_k P_l, \quad (A.4)$$

that includes conventional Hook relation, chemical stresses, flexoelectric and electrostriction contributions. Since the static equation of mechanical equilibrium, $\partial \sigma_{ij}/\partial x_j = 0$, should be valid, the equation (A.4) transforms into Lame-type equation for elastic displacement $U_i$

$$\frac{\partial \sigma_{mn}}{\partial x_m} = c_{ijmn}\frac{\partial}{\partial x_m}\left(u_{ij} - W_{ij}\delta N_d + F_{ijkl}\frac{\partial P_l}{\partial x_k} - Q_{ijkl}P_k P_l\right) = 0. \quad (A.5a)$$

Equivalents form of Eq.(A.5a) is

$$c_{ijkl}\frac{\partial^2 U_l}{\partial x_j \partial x_k} = \frac{\partial}{\partial x_j}\left(\sigma_{ij}^W - f_{ijkl}\frac{\partial P_l}{\partial x_k} + q_{ijkl}P_k P_l\right) \quad (A.5b)$$

Where $c_{ijkl}$ are elastic stiffness, random Vegard stresses $\sigma_{mn}^W = c_{ijmn}W_{ij}\delta N_d$, $q_{mnkl} = c_{ijmn}Q_{ijkl}$ is electrostriction stress tensor and $f_{mnkl} = c_{ijmn}F_{ijkl}$ is the flexocoupling stress tensor.

In order to derive expression for the linear generalized susceptibility and correlation function, let us linearize Eqs.(A.2) for polarization and Eq. (A.5) for the elastic displacement.

In Fourier space of spatial wave vector **k** the linearized solutions have the form:

$$\delta P_i(\mathbf{r}) = \frac{1}{(2\pi)^{3/2}}\int_{-\infty}^{\infty} d\mathbf{k}\,\delta\widetilde{P}_i \exp(-ik_j x_j), \qquad \delta U_i(\mathbf{r}) = \frac{1}{(2\pi)^{3/2}}\int_{-\infty}^{\infty} d\mathbf{k}\,\delta\widetilde{U}_i \exp(-ik_j x_j), \quad (A.6a)$$

Next the electric field in Eq.(A.2) can be represented as $E_i = \delta E_i + E_i^{ext}$, where the local fluctuations of depolarization field

$$\delta E_i(\mathbf{r}) = \frac{1}{(2\pi)^{3/2}}\int_{-\infty}^{\infty} d\mathbf{k}\,\delta\widetilde{E}_j^{dep} \exp(-ik_j x_j), \quad (A.6b)$$

can be estimated in Debye approximation as described below. The Fourier image of the Poisson equation (A.3) for the depolarization field determination in Debye approximation acquires the form

$$-k^2\widetilde{\varphi} \approx -\frac{ik_i\delta\widetilde{P}_i}{\varepsilon_b\varepsilon_0} + \frac{\widetilde{\varphi}}{R_d^2}, \quad (A.7)$$

and its solution is

$$\delta\widetilde{E}_j \approx \frac{-k_m k_j \delta\widetilde{P}_m}{\varepsilon_b\varepsilon_0(k^2 + R_d^{-2})}, \quad (A.8)$$

where $R_d$ is the screening radius that depends on temperature and singly-ionized defect concentration as $R_d = \sqrt{\dfrac{\varepsilon_b\varepsilon_0 k_B T}{2e^2 \overline{N}_d}}$.

Allowing for Eqs.(A.8) the Fourier representations of the linearized equations (A.5b) and (A.2) have the form for Fourier amplitudes:

$$c_{ijkl}k_l k_j \delta\widetilde{U}_k + ik_j \widetilde{\sigma}^W_{ij} + f_{ijml}k_j k_m \delta\widetilde{P}_l = 0, \qquad (A.9a)$$

$$\left(\alpha\delta_{ij} + \frac{k_i k_j}{\varepsilon_b \varepsilon_0 (k^2 + R_d^{-2})} + g_{imjl}k_m k_l\right)\delta\widetilde{P}_j + f_{mnli}k_l k_n \delta\widetilde{U}_m = \widetilde{E}_i^{ext}. \qquad (A.9b)$$

The strain tensor components are related with displacement components from Eq.(A.9a) as $\delta\widetilde{u}_{mn} = -i(k_n \delta\widetilde{U}_m + k_m \delta\widetilde{U}_n)/2$.

The Fourier representations in the spatial **k** and frequency ω domain of the linearized solution for polarization and strain fluctuations have the form:

$$\delta\widetilde{P}_j(\mathbf{k}) = \left[\widetilde{E}_i^{ext} + ik_{j'} S_{mi'}(\mathbf{k}) f_{mnli}k_l k_n \widetilde{\sigma}^W_{i'j'}\right]\widetilde{\chi}_{ij}(\mathbf{k}), \qquad (A.10a)$$

$$\delta\widetilde{U}_k(\mathbf{k}) = -ik_j \widetilde{\sigma}^W_{ij} S_{ik}(\mathbf{k}) + S_{ik}(\mathbf{k})\widetilde{\chi}_{sl}(\mathbf{k}) f_{ijml}k_j k_m \left[\widetilde{E}_s^{ext} + if_{qnps}k_p k_n k_{j'} S_{qi'}(\mathbf{k})\widetilde{\sigma}^W_{ij}\right]. \qquad (A.10b)$$

Since the harmonic approach (5) is applicable for small wave vector **k**, consideration of the problem for higher **k** values requires including of the anharmonicity and higher gradient terms.

Generalized susceptibility $\widetilde{\chi}_{ij}(\mathbf{k})$, that is in fact correlation function, and elastic function $S_{ir}(\mathbf{k})$ included in Eqs.(A.10) are given by expressions:

$$\widetilde{\chi}_{ij}^{-1}(\mathbf{k}) = \beta_{ij}(\mathbf{k}) + \Theta_{ipjl}(\mathbf{k}), \qquad S_{ik}^{-1}(\mathbf{k}) = c_{ijkl}k_l k_j. \qquad (A.11a)$$

Here the linear dynamic stiffness is affected by depolarization effect as

$$\beta_{ij}(\mathbf{k}) = \alpha\delta_{ij} + \frac{k_i k_j}{\varepsilon_b \varepsilon_0 (k^2 + R_d^{-2})}, \qquad (A.11b)$$

$$\Theta_{ipjl}(\mathbf{k}) = g_{ipjl}k_p k_l - f_{mnli}k_n k_l f_{i'j'pj}k_{j'}k_p S_{mi'}(\mathbf{k}). \qquad (A.11c)$$

Substitution of Eqs.(A.11) into Eq.(A.10a) allowing for the smallness of the flexoelectric coupling and Vegard strains, and condition $\widetilde{E}_i^{ext} = 0$ yields to the approximation

$$\delta\widetilde{P}_j(\mathbf{k}) \approx if_{mnli}k_l k_n k_{j'} S_{mi'}(\mathbf{k})\widetilde{\sigma}^W_{i'j'}\widetilde{\chi}_{ij}(\mathbf{k}), \qquad (A.12a)$$

$$\widetilde{\chi}_{ij}^{-1}(\mathbf{k}) \approx \alpha\delta_{ij} + \frac{k_i k_j}{\varepsilon_b \varepsilon_0 (k^2 + R_d^{-2})} + g_{ipjl}k_p k_l, \qquad S_{ik}^{-1}(\mathbf{k}) = c_{ijkl}k_l k_j. \qquad (A.12b)$$

The inverse matrices of the Green tensor $\widetilde{\chi}_{ij}(\mathbf{k})$ is in fact correlation function or generalized susceptibility $\left.\frac{\partial \widetilde{P}_i(\mathbf{k})}{\partial E_j^{ext}}\right|_{E_j^{ext}\to 0} \equiv \widetilde{\chi}_{ij}(\mathbf{k})$. Order parameter correlation function is related with the Green tensor via Callen-Welton fluctuation-dissipation theorem, and corresponding static correlations radius can be determined from direct matrix $\widetilde{\chi}_{ij}(\mathbf{k})$.

Further analytical estimations in Eqs.(A.12) are possible only for the simplest spherically symmetric dilatation center in an isotropic surrounding (while more realistically it is cubic). For the case the components of gradient tensor, elastic function and dielectric susceptibility are

$$g_{ipil}k_pk_l = (g_{11} - g_{44})k_ik_j + \delta_{ij}g_{44}k^2, \qquad g_{12} = (g_{11} - g_{44})/2, \qquad (A.13a)$$

$$S_{ij}(\mathbf{k}) = \frac{1}{c_{44}k^2}\left(\delta_{ij} - \frac{k_ik_j}{k^2}\right) + \frac{1}{c_{11}}\frac{k_ik_j}{k^4}, \qquad (A.13b)$$

$$\tilde{\chi}_{ij}(\mathbf{k}) \approx \left(\delta_{ij} - \frac{k_ik_j}{k^2}\right)\frac{1}{\alpha + g_{44}k^2} + \frac{k_ik_j}{k^2}\frac{1}{\alpha + g_{11}k^2 + k^2/(\varepsilon_b\varepsilon_0(k^2 + R_d^{-2}))}. \qquad (A.13c)$$

Using above expressions it makes sense to estimate **k**-spectra Eq.(A.12a) is spherical coordinates:

$$\delta\vec{\tilde{P}}(\mathbf{k}) \sim \frac{-if_{11}\mathbf{k}W_{eff}\delta\tilde{N}_d(\mathbf{k})}{\alpha + g_{11}k^2 + k^2/(\varepsilon_b\varepsilon_0(k^2 + R_d^{-2}))} \qquad (A.14)$$

Where the effective Vegard coefficient $W^{eff} = W_{11}\left(1 + 2\frac{c_{12}}{c_{11}}\right)$ is introduced. The second term $g_{11}k^2$ in the denominator originated from the polarization gradient term, and the third term $\frac{k^2}{\varepsilon_b\varepsilon_0(k^2 + R_d^{-2})}$ originated from depolarization field $\delta\tilde{E}_j \approx \frac{-k_mk_j\delta\tilde{P}_m}{\varepsilon_b\varepsilon_0(k^2 + R_d^{-2})}$, which were estimated in Debye approximation, and $R_d$ is a Debye screening radius. The expression in **r**-space is

$$\delta\vec{P}(\mathbf{r}) \cong \frac{1}{(2\pi)^{3/2}}\int_{-\infty}^{\infty}d\mathbf{k}\frac{-if_{11}W_{eff}\mathbf{k}\delta\tilde{N}_d(\mathbf{k})\exp(-i\mathbf{k}\mathbf{r})}{\alpha + g_{11}k^2 + k^2/(\varepsilon_b\varepsilon_0(k^2 + R_d^{-2}))} \qquad (A.15)$$

In order to find a solution in **r**-space one should average Eq.(A.15) with the spectra of the distribution function of vacancies $\delta\tilde{N}_d(\mathbf{k})$ and then make an inverse Fourier transformation. Since $\langle\delta\tilde{N}_d(\mathbf{r})\rangle = 0$ the average value $\langle\delta\vec{P}(\mathbf{r})\rangle$ is zero as anticipated, and the expressions for averaged mean squire value $\sqrt{\langle\delta\vec{P}^2(\mathbf{r})\rangle}$ can be derived after analyzing the correlation function $\langle\delta\vec{P}(\mathbf{r})\delta\vec{P}^*(\mathbf{r}')\rangle$:

$$\langle\delta\vec{P}(\mathbf{r})\delta\vec{P}^*(\mathbf{r}')\rangle \cong \frac{1}{(2\pi)^3}\int_{-\infty}^{\infty}d\mathbf{k}'\int_{-\infty}^{\infty}d\mathbf{k}\frac{(f_{11}W_{eff})^2\mathbf{k}\mathbf{k}'\langle\delta\tilde{N}_d(\mathbf{k})\delta\tilde{N}_d(\mathbf{k}')\rangle\exp(-i\mathbf{k}\mathbf{r} + i\mathbf{k}'\mathbf{r}')}{\left(\alpha + g_{11}k'^2 + \frac{k'^2}{\varepsilon_b\varepsilon_0(k'^2 + R_d^{-2})}\right)\left(\alpha + g_{11}k^2 + \frac{k^2}{\varepsilon_b\varepsilon_0(k^2 + R_d^{-2})}\right)}$$

(A.16)

Assuming that the **k**-spectra of vacancies distribution has the form

$$\langle\delta\tilde{N}_d(\mathbf{k})\delta\tilde{N}_d(\mathbf{k}')\rangle = (2\pi)^3 \delta(\mathbf{k}-\mathbf{k}')n_d^2 f_d(\mathbf{k}), \qquad (A.17)$$

the correlation function becomes:

$$\langle\delta\vec{P}(\mathbf{r})\delta\vec{P}^*(\mathbf{r}')\rangle \cong (f_{11}W_{eff}n_d)^2 \int_{-\infty}^{\infty} d\mathbf{k}\, \frac{k^2 f_d(\mathbf{k})\exp(-i\mathbf{k}(\mathbf{r}-\mathbf{r}'))}{\left(\alpha + g_{11}k^2 + k^2/\varepsilon_b\varepsilon_0(k^2+R_d^{-2})\right)^2} \qquad (A.18)$$

Autocorrelation function ($\mathbf{r}=\mathbf{r}'$) has the form

$$\langle\delta\vec{P}^2(\mathbf{r})\rangle \cong \int_{-\infty}^{\infty} d\mathbf{k}\, \frac{(f_{11}W_{eff}n_d)^2 k^2 f_d(\mathbf{k})}{\left(\alpha + g_{11}k^2 + k^2/\varepsilon_b\varepsilon_0(k^2+R_d^{-2})\right)^2} \qquad (A.19)$$

Let us assume a sharp enough Gaussian distribution of the vacancies

$$f_d(\mathbf{k}) \cong \frac{1}{(\sqrt{2\pi}\Delta)^3}\exp\left(-\frac{(\mathbf{k}-\mathbf{k}_c)^2}{2\Delta^2}\right), \qquad (A.20)$$

with the halfwidth $\Delta \ll R_d^{-2}$ and the most probable value $k_c \cong \xi\cdot\sqrt[3]{N_d}$. Since $k_c$ value defines the period of long-range correlations the dimensionless parameter $\xi$ should be in order of unity and $k_c \gg \frac{1}{r_0}$. Using Eq.(A.20) the integral in Eq.(A.19) can be estimated using Laplace method:

$$\langle\delta\vec{P}^2(\mathbf{r})\rangle \cong \frac{(f_{11}W_{eff}n_d k_c)^2}{\left(\alpha + g_{11}k_c^2 + \dfrac{k_c^2 R_d^2}{\varepsilon_b\varepsilon_0(1+k_c^2 R_d^2)}\right)^2} \qquad (A.21)$$

The expression for the autocorrelation of depolarization random electric field $\langle\delta\vec{E}^2(\mathbf{r})\rangle$ can be derived in a similar way using Eq.(A.8):

$$\langle\delta\vec{E}^2(\mathbf{r})\rangle \cong \frac{k_c^2 R_d^4 \langle(\mathbf{k}_c\delta\vec{P})^2\rangle}{(\varepsilon_b\varepsilon_0)^2(1+k_c^2 R_d^2)^2}. \qquad (A.22)$$

Allowing for the spherical symmetry of defect center the latter expressions

$$\sqrt{\langle\delta\vec{P}^2(\mathbf{r})\rangle} \cong \frac{|f_{11}W_{eff}k_c|\cdot n_d}{\left|\alpha_T(T-T_C^*) + g_{11}k_c^2 + \dfrac{k_c^2 R_d^2}{\varepsilon_b\varepsilon_0(1+k_c^2 R_d^2)}\right|}, \qquad (A.23a)$$

$$\sqrt{\langle\delta\vec{E}^2(\mathbf{r})\rangle} \cong \frac{k_c^2 R_d^2 |f_{11}W_{eff}k_c|\cdot n_d}{\varepsilon_b\varepsilon_0(1+k_c^2 R_d^2)\left|\alpha_T(T-T_C^*) + g_{11}k_c^2 + \dfrac{k_c^2 R_d^2}{\varepsilon_b\varepsilon_0(1+k_c^2 R_d^2)}\right|}. \qquad (A.23b)$$

Where $n_d \equiv \sqrt{\langle\delta N_d^2\rangle}$ allowing for Eq.(A.17) and (A.20).

## APPENDIX B. Auxiliary derivations and schemes

### B.1. Derivation of Eq.(6)

Using ergodic hypothesis the averaging in Eq.(5) over the defects distribution function is reducing to the averaging over the defect partial volume, $V = \frac{4\pi}{3}R^3$, and gives the following expression:

$$\left\langle \alpha_{kl}^{R}(T,\overline{N}_d) \right\rangle \approx a_{kl}(T) - 2Q_{ijkl}\left\langle \sigma_{ij}^{W}(\mathbf{r}) \right\rangle = a_{kl}(T) - 2Q_{ijkl}\tilde{W}_{ij} \frac{3}{4\pi R^3} \int_0^{2\pi} d\varphi \int_0^{\theta} \sin\theta d\theta \int_{r_0}^{R} \frac{r^2 dr}{2\pi r^3}$$
$$= a_{kl}(T) - 4Q_{ijkl}\tilde{W}_{ij} \frac{3}{4\pi R^3} \ln\left(\frac{R}{r_0}\right) \equiv a_{kl}(T) - \frac{4}{3}Q_{ijkl}\tilde{W}_{ij}\overline{N}_d \ln\left(\frac{3}{4\pi \overline{N}_d r_0^3}\right)$$ (B.1)

where $\tilde{W}_{ij} = c_{ijkl}W_{kl}$ and $c_{ijkl}$ is the elastic stiffness tensor.

### B.2. Possible role of the intrinsic charge of the vacancies (if any).

If one put formally into unoccupied point of $Ti^{4+}$ charges $+4e$ and $-4e$ the charge $+4e$ makes ideal lattice and $-4e$ corresponds to defect in it. The similar formal operation with the addition to $Ti^{3+}$ position $+e$ and $-e$ leads to the appearance into ideal lattice defect – $4e + (Ti^{3+} + e)$, that is dipole $d_1 = 4es$, where $s$ is $Ti^{3+}$ off-central shift. An illustration of the oxygen vacancy related defect configurations in a tetragonal perovskite lattice structure is shown in **Fig. S1**.

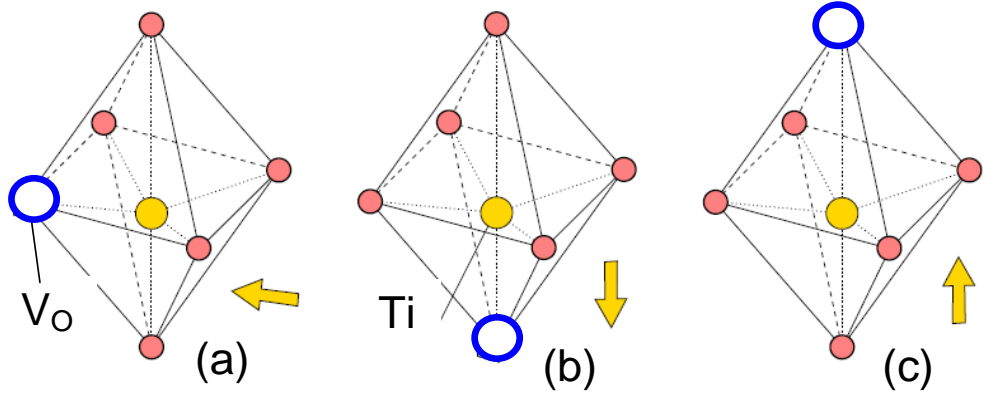

**Figure S1.** Oxygen vacancy related defect configurations in a tetragonal perovskite lattice structure. **(a)**–**(c)** Possible configurations for the dipole-acceptor associate with the defect dipole aligned in different directions (shown by arrows). (Adapted from [Erhart, Paul, and Karsten Albe. "Dopants and dopant–vacancy complexes in tetragonal lead titanate: A systematic first principles study." *Computational Materials Science* 103 (2015): 224-230]).